\documentclass[]{spie}  
\usepackage[]{graphicx}

\title{Japanese Virtual Observatory (JVO) as an advanced astronomical research enviroment} 

\author{Yuji Shirasaki\supit{a}, 
Masahiro Tanaka\supit{a}, 
Satoshi Kawanomoto\supit{a},
Satoshi Honda\supit{a}, 
Masatoshi Ohishi\supit{a},
Yoshihiko Mizumoto\supit{a},  
Naoki Yasuda\supit{b},  
Yoshifumi Masunaga\supit{c},  
Yasuhide Ishihara\supit{d},  
Jumpei Tsutsumi\supit{d},  
Hiroyuki Nakamoto\supit{e},
Yuusuke Kobayashi\supit{e} and
Michito Sakamoto\supit{e}
\skiplinehalf
\supit{a}National Astronomical Observatory of Japan, 2-21-1 Osawa, Mitaka Tokyo, 181-8588 Japan\\
\supit{b}University of Tokyo, 5-1-5 Kashiwa-no-Ha, Kashiwa Chiba, 277-8582 Japan\\
\supit{c}Ochanomizu Univerisity, 2-1-1 Otsuka Bunkyo-ku, Tokyo, 112-8610 Japan\\
\supit{d}Fujitsu Ltd., 4-1-1 Kamikodanaka Nakahara-ku, Kawasaki, 211-8588 Japan\\
\supit{e}Systems Engineering Consultants Co. Ltd., 22-4 Sakuraoka-cho Shibuya-ku, Tokyo, 150-0031 Japan\\
}

\authorinfo{Further author information: (Send correspondence to Y.S.)\\
Y.S.: E-mail: yuji.shirasaki@nao.ac.jp, 
Telephone: 81 422 34 3579\\
}

\begin{document} 
\maketitle 

\begin{abstract}
We present the design and implementation of the Japanese Virtual Observatory 
(JVO) system. JVO is a portal site to various kinds of astronomical resources
distributed all over the world. We have developed five components for constructing
the portal: (1) registry, (2) data service, (3) workflow system, (4) data 
analysis service (5) portal GUI. Registry services are used for publishing and 
searching data services in the VO, and they are constructed using an OAI-PMH 
metadata harvesting protocol and a SOAP web service protocol so that VO standard
architecture is applied. Data services are developed based on the
Astronomical Data Query Language (ADQL) which is an international VO
standard and an extension of the standard SQL. The toolkit for building
the ADQL-based service is released to the public 
on the JVO web site. The toolkit also provides the protocol translation from 
a Simple Image Access Protocol (SIAP) to ADQL protocol, so that both the VO 
standard service can be constructed using our toolkit. In order to federate 
the distributed databases and analysis services, we have designed a workflow
language which is described in XML and developed execution system of the workflow.
We have succeeded to connect to a hundred of data resources of the world as 
of April 2006. We have applied this system to the study of QSO environment
by federating a QSO database, a Subaru Suprim-Cam database, and some analysis 
services such a SExtractor and HyperZ web services. These experiences are described
is this paper.

\end{abstract}



\section{INTRODUCTION}
\label{sect:intro}  

The recent progress of the information technology represented by a Web/Grid 
service, workflow system, single-sign-on system enables constructing a
global astronomical research environment over the internet.
Virtual Observatories (VOs) are emerging research environment for
astronomy, and 16 countries and regions have funded to develop their
VOs based on international standard protocols for interoperability. 
National Astronomical Observatory of Japan (NAOJ) started its VO project
(Japanese Virtual Observatory -- JVO) in 2002, and developed its 
prototypes.~\cite{Ohishi2004} \cite{Tanaka2004}
\cite{Tanaka2005} \cite{Ohishi2006} \cite{Shirasaki2006}
At the early stage of the project our efforts has been mainly focused 
on the database federation.
Now that we have succeeded to interoperate the VO data services in 2004,
we have started to integrate an analysis system to the JVO prototype
in 2005.
Importance of the multi-wavelength study is rising in the recent years, 
however, difficulty for doing such study remains.
One reason is that, for each data set, one needs to learn how to reduce 
and analyze the data, and even needs to know where the analysis tools 
are available.
To overcome such difficulty, seamless access to not only the data but also
to the analysis tools needs to be achieved.
From that view point, we have developed workflow system which federates 
both of the data and analysis service over the Internet.
To prove usefulness of the VO for astronomical research, we needs to 
demonstrate how is the research achieved efficiently using the VO system.
We, therefore, applied our JVO system to the cosmic string
search~\cite{Shirasaki2004} and the QSO environment
study~\cite{Shirasaki2006}, successfully demonstrated how efficient is
the VO system.
%


\section{OVERVIEW OF THE JVO SYSTEM} 

\begin{figure}
   \begin{center}
   \begin{tabular}{c}
   \includegraphics[height=7cm]{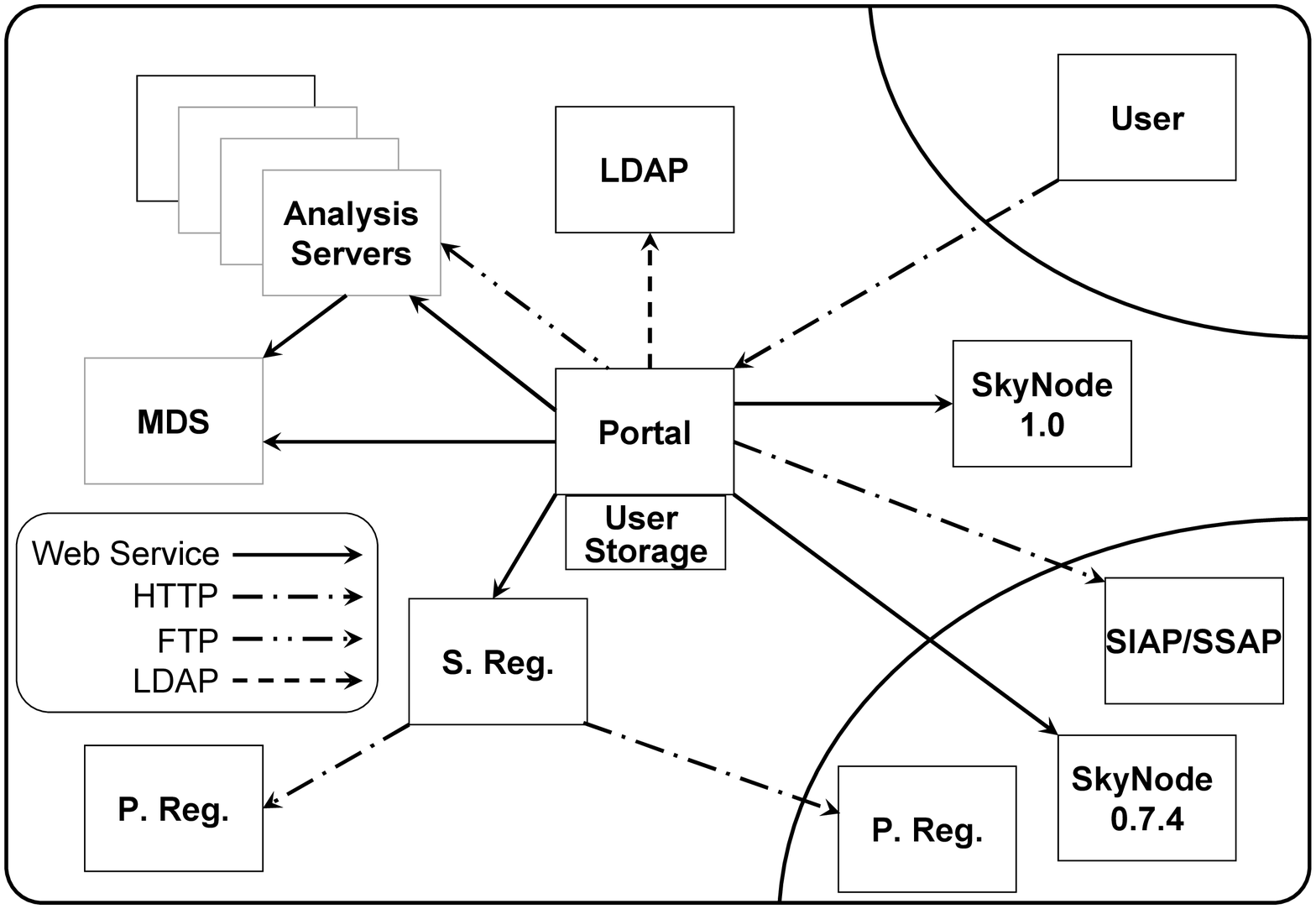}
   \end{tabular}
   \end{center}
   \caption[portal] 
   { \label{fig:portal-overview} 
   The diagram of the current JVO system.
   }
\end{figure} 

The JVO portal service provides a seamless and homogeneous access 
to the astronomical data service.
Accessing the portal, one can get various kinds of data (object catalog,
image and spectrum data, and so on), which are produced and managed at
a various level of organizations (an observatory, a data center, 
a science project or a researcher).
The JVO portal will also provide an analysis environment utilizing
the grid computing system, where one can visualize data at various
aspects (XY plot, contour plot, color image, ...), calculate 
physical quantity from the data (source flux, coordinate, morphological
parameter, photometric redshift, ...), and reduce the data especially
of Subaru with the user supplied parameters.
Fig.~\ref{fig:portal-overview} shows a diagram of the current JVO
system.
The portal server interacts with the other local servers, to authenticate
a user (LDAP server), to find a data resource in the VO (Searchable
Registry) and to submit an analysis job (MDS server).
The protocol mostly used for the communication with each service is an
HTTP, especially a web service.
The HTTP protocol is widely used in the internet world and is a
relatively firewall-friendly protocol, so the system is easily extended
to the larger-scale network system.

\begin{figure}
   \begin{center}
   \begin{tabular}{c}
   \includegraphics[width=1.0\textwidth]{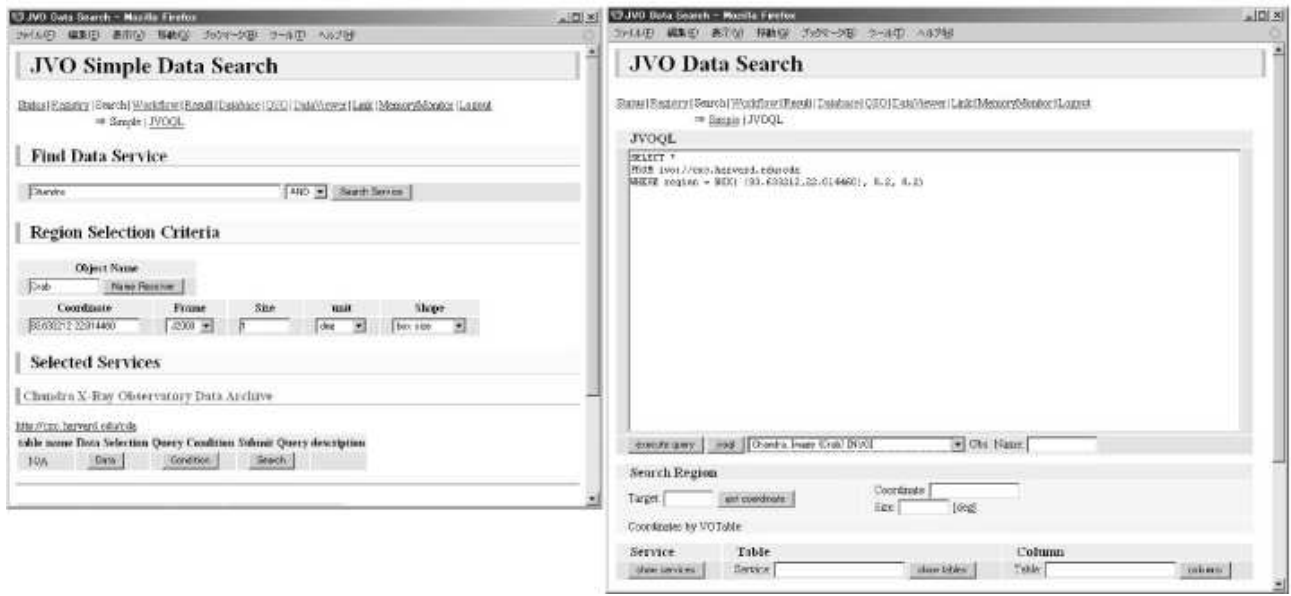}
   \end{tabular}
   \end{center}
   \caption[serach-gui] 
   { \label{fig:search-gui} JVO data search GUI. The left figure shows 
   a search form where one can specify the service and criterion of
   the search region. The right figure shows a SQL based search form
   where one can write a JVOQL2 to send a query for multiple
   databases.}
\end{figure}

The user sends a request through his/her own web browser in various
formats (Fig.~\ref{fig:search-gui}), web form based request, SQL based
data query, and workflow description language constructed by a
graphical/command user interface.
According to the request, the portal service creates a workflow to resolve
the service location with the searchable registry and the Monitoring
and Discovery service (MDS), submit a query or job to the VO service of
local and of the other VO (right bottom of the
Fig.~\ref{fig:portal-overview}), and retrieve and store the data on the
user's storage area.
The JVO portal service is not yet open to the public, but will become 
public soon in this year 2006 or early of 2007.
It will not require user registration for using the basic and limited 
functionally, such as a data resource search on the registry,
a data search which utilizes the skynode and SIA/SSA data service and does 
not consume much resource on the portal system, and a short-lived
analysis service.
The user registration will be required to use the full functionality
of the portal service (a big query, workflow, a long-lived job).
%

\section{SKYNODE SERVICE}

We are developing a SkyNode toolkit for building a VO compliant data service. 
Skynode is a VO standard data access interface~\cite{Budavari2004}.
A query is described in SQL-like syntax with astronomical extensions, and is 
transfered as a SOAP message over HTTP protocol.
The aim of this development is to reduce the time and effort for the
data providers to spend for implementing a skynode interface on their
database.
Reducing the cost for introducing the VO standard is the most important
factor for widely deploying the VO standard services.
First primitive version was released in May 2005 at the JVO web site
\footnote{http://jvo.nao.ac.jp/download/},
and update versions are also available at the same URL.
Fig.~\ref{fig:skynode} shows the architecture of the JVO SkyNode.
The toolkit can be used for building a service which supports Simple Image
Access Protocol (SIAP) and Simple Spectrum Access Protocol (SSAP), which
are also defined as VO standards.
The SIAP/SSAP are based on a simple URL parameter query and it is
translated to the SQL internally.
For an example, a query like
\begin{verbatim}
   http://jvo.nao.ac.jp/search?POS=34,20&SIZE=1.0&PARAM1=VALUE1&...
\end{verbatim}
is translated to the following SQL:
\begin{verbatim}
   SELECT * FROM table WHERE Region('BOX 34 20 1.0') and param1=value1 and ...
\end{verbatim}
The region condition specified by \verb|POS| and \verb|SIZE| parameters is 
translated to an ADQL region expression, and the other conditions are just
connected by a logical operator \verb|AND|.
The ADQL is translated to an SQL for backend database management system and
the interface to the backend database is provided by Java DataBase Connectivity 
(JDBC).
The translation is performed to be in conformity with the SQL92 syntax as 
mush as possible, so this toolkit can be applied to most of the DBMSs.
Exception is that ADQL \verb|TOP| syntax is not SQL92 conformance, so
there is a diversity among the DBMSs.
Such a difference of the syntax is managed by preparing a plug-in Java
translator for each DBMS.
Currently we are testing the functionality of the toolkit on PostgreSQL
and MySQL, the other DBMS will be tested on a future release.
The ADQL \verb|Region| and \verb|Xmatch| expression is converted to a SQL92 syntax
by using table join construct, so there is no DBMS dependence for the two
ADQL specific expressions.
The region search is performed by using HTM index~\cite{Kunszt2001}.
For a table that has spatial coordinate columns, HTM index is stored on the
database in advance.
The region specified in the ADQL can be represented by a series of HTM index 
ranges, so the lower (\verb|htm_low|) and upper (\verb|htm_upp|)
boundary of each HTM index range are stored on a table.
The region search can be performed by taking a join between the HTM range table 
(t2), which is supplied by the search region condition, and the HTM
index table (t1) of the queried table.
The join condition is \verb|t1.htm between t2.htm_low and t2.htm_upp|.
The toolkit also provides cross-match functionality, that is, selecting data
for multiple regions specified in a VOTable.
Cross-match is equivalent to a query where union of multiple regions is specified
as a search condition, so it can be performed by the same procedure as the single
region search.
Data resources currently available from JVO are: (1) Subaru Deep field survey 
catalogs and images, (2) Subaru Suprime-Cam Open Data Archive, (3) QSO catalog 
compiled by Veron et al. and copied from VizieR, (4) SDSS DR2. (5) TWOMASS
Catalog.
%
\begin{figure}[t]
   \begin{center}
   \includegraphics[height=7cm]{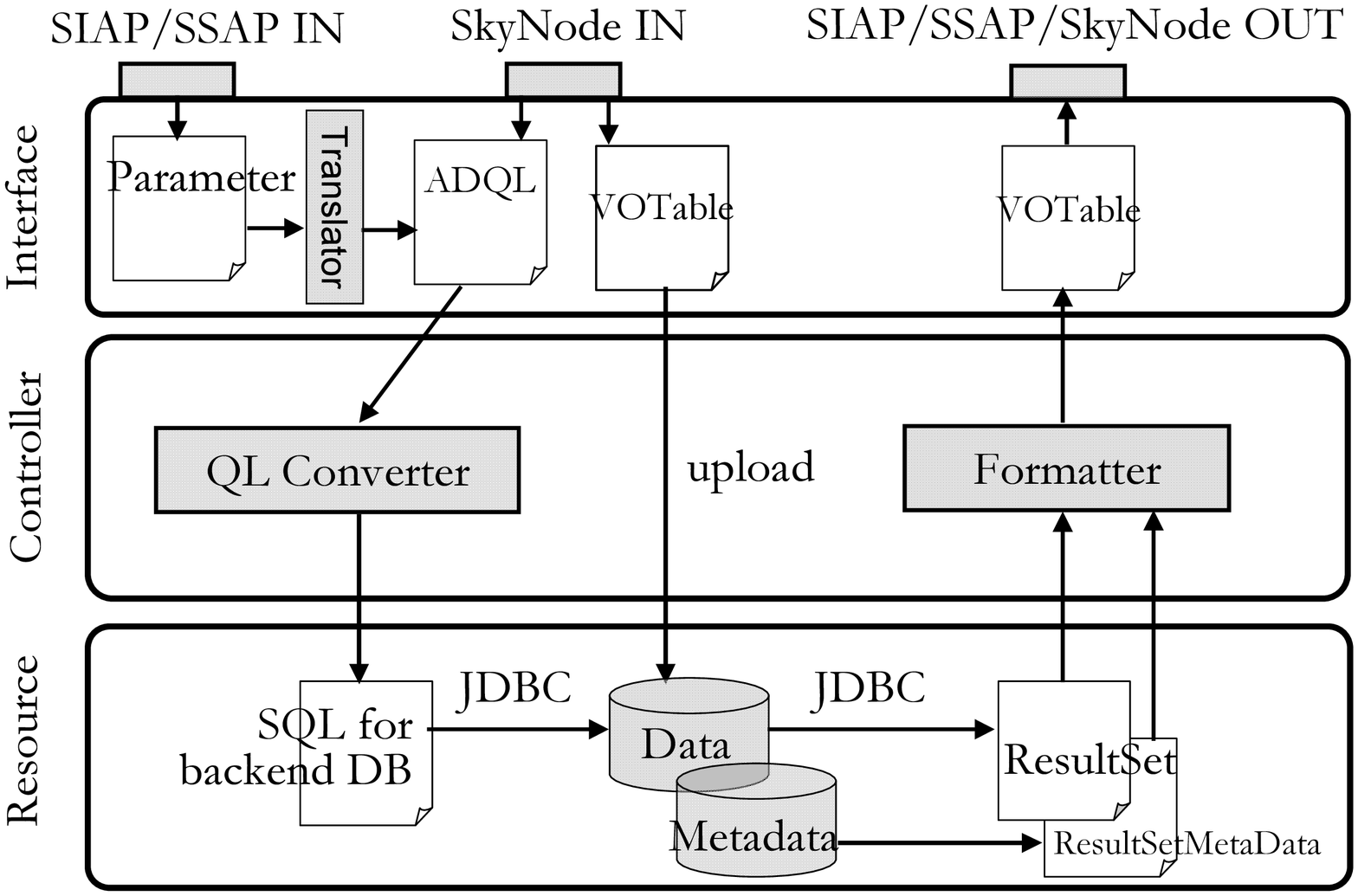}
   \end{center}
   \label{fig:skynode}
   \caption[skynode] 
   { \label{fig:skynode} 
      The architecture of the JVO skynode.
   }
\end{figure} 



\section{REGISTRY} 

We have constructed a publishing registry and a searchable registry.
The role of the publishing registry is to expose any kind of 
resource metadata to the VO world.
The OAI-PMH
protocol\footnote{http://www.openarchives.org/OAI/openarchivesprotocol.html},
which is defined originally for exchanging the metadata of an electric
document, is used for retrieving the metadata from the publishing registry.
The publishing registry was made by using the NVO
software~\cite{Williamson2004}, and which was slightly modified to adapt
the recent VOResource schema.

\begin{figure}[h]
   \begin{center}
   \begin{tabular}{c}
   \includegraphics[width=0.9\textwidth]{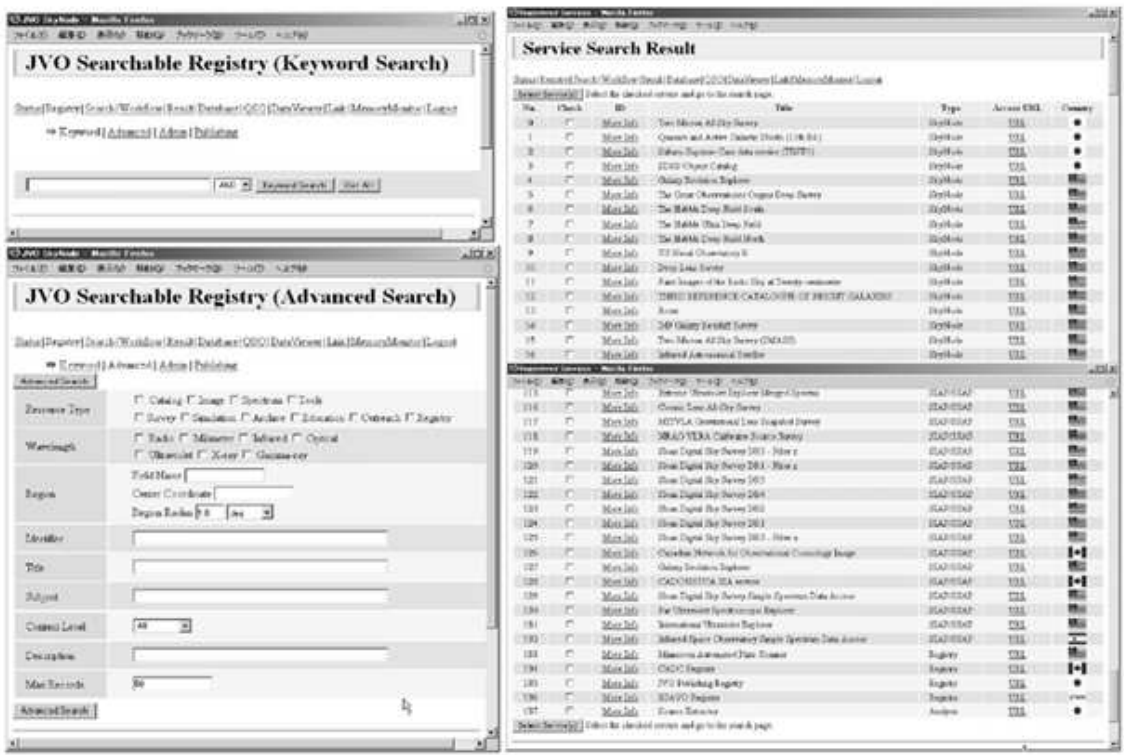}
   \end{tabular}
   \end{center}
   \caption[registry] 
   { \label{fig:registry} A graphical user interface for the JVO
 searchable registry. The left top figure shows the keyword based
 search page, the left bottom one shows the advanced search page where
 one can specify keywords for each resource attribute. The right figure
 shows a search result for all the registered resources.}
\end{figure}

JVO searchable registry is constructed of a native XML DB, 
Karearea\textcopyright (SEC), the tomcat web server, and the AXIS SOAP 
message handler.
The searchable registry collects metadata from publishing registries 
not only of JVO but also of the other VO projects.
Currently we are collecting from STSCI, NCSA, ESAC, CDS, HEASARC and JVO.
At the time of implementation, a lot of metadata collected from these
publishing registries were not valid XML documents, so we needed to modify 
the resource metadata schema so that we accept such an invalid metadata.
Currently only the metadata of which the resource type is one of
Registry, OpenSkynode, SIAP/SSAP and JVOAnalysis.
Three VO standard interfaces, the keyword-based search, the 
identifier-based search, and the ADQL search are implemented.
A user an access to the the searchable registry on his/her web browser
as shown in Fig.~\ref{fig:registry}.
JVO searchable registry also collects table metadata from skynodes.
These data are used for resolving the service endpoint URL from the
SQL table name, and for assisting a user to make a query.
JVO searchable registry have extra interfaces to access to the table metadata.
It is also possible to register a resource metadata directly without
using the publishing registry.

The performance of the metadata harvesting was measured as shown in
table~\ref{tbl:registry_performance}.
The total time for completing the harvesting was 36 minutes 52 seconds,
70 \% of the time was spent for waiting the connection time out for 
servers being down.
This will be a serious bottleneck, so we need to resolve this issue
by, for an example, multi-thread approach.
\begin{table}
\caption{Processing time budget for each metadata harvesting.}
\label{tbl:registry_performance}
\begin{center}
\begin{tabular}{|p{0.3\textwidth}|r|p{0.5\textwidth}|}\hline
\multicolumn{1}{|c|}{item}
&\multicolumn{1}{|c|}{time}
& \multicolumn{1}{|c|}{comment} 
\\ \hline
total time consumed for retrieving metadata from live servers
&  6 min 2 sec & 
543 HTTP transactions, 455 ms for each connection
 establishment -- 4 min 7 sec in total
\\ \hline
total time consumed for retrieving metadata from servers being down
& 26 min 15 sec & 
  7 HTTP transactions 
\\ \hline
total time consumed for registering the data to the XML DB
& 4 min 35 sec & includes time for parsing the OAI-PMH and SOAP
  messages and updating the XML DB.
\\ \hline
total
& 36 min 52 sec & total data size = 12.3 MB (0.5 MB from publishing
 registries and 11.8 MB from SkyNode) \\
\hline
\end{tabular}
\end{center}
\end{table}

\section{JVO QUERY LANGUAGE 2}

We defined the second version of JVO query language (JVOQL-2) to
describe a query to the distributed multiple databases in a single
statement.
The JVOQL was a language dedicated for the JVO search engine and a JVO
advanced user, but now it is also a prototype language for the VO query
language as an international standard.
The first version of JVOQL (JVOQL-1) was discussed elsewhere~\cite{Tanaka2004}.
The main difference from the version one is the way to describe a query
to retrieve an image and table name specification.
According to the specification of JVOQL-1, the region of the image to be
extracted from an archive is specified in a functional expression at the
selection list.
In JVOQL-2 new data types that represent spatial coordinates and a
region are introduced to specify a search region in a same manner as
specified for the ordinal data type.
So the SQL expression of the image data query is now simple and similar
to that of an ordinal query.
The regional coverage of the retrieved image can be specified at the
\verb|WHERE| clause and not at the selection list, and the role
separation of the selection list and \verb|WHERE| clause is now clearer
than the case of JVOQL-1.

In order to specify a table uniquely in the VO, resource identifier is
followed by a colon and a table name.
The following is an example for searching on a table \verb|image|
provided on a service of which the identifier is
\verb|ivo://jvo/subaru/spcam| with region condition specified in the
\verb|WHERE| clause.
\begin{verbatim}
    SELECT access_ref
    FROM   ivo://jvo/subaru/spcam:image
    WHERE  region = Box((230, +10), 1.0, 1.0)
\end{verbatim}
The column \verb|access_ref| returns access URLs of the searched image.
The column \verb|region| corresponds to the region of the image that will
be retrieved from the URL and its data type is \verb|spatialRegion|.
In the JVO skynode toolkit, the column of an extended data type is
implemented as a function type column.
In the above example, the identifier \verb|ivo://jvo/subaru/spcam| may be
expressed by a dot and colon notation like \verb|jvo:subaru.spcam|.
The above search region can be specified in a VO standard form like 
\verb|Region('Box 230 +10 1.0 1.0')|.
It is also possible to describe a query to the SIAP service by specifying the
service identifier at the \verb|FROM| clause and omit a colon and table name.
\begin{verbatim}
    SELECT * 
    FROM   ivo://cxc.harvard.edu/cda
    WHERE  Region('BOX 83.633 22.014 0.2, 0.2')
\end{verbatim}
Table join between the object catalog and image data table can be described
like:
\begin{verbatim}
    SELECT  qso.*, img.* 
    FROM    ivo://jvo/vizier/VII/235:qso_veron_2003 qso,
            ivo://jvo/subaru/spcam:spcam_mos_view AS img
    WHERE   Point(qso.raj2000, qso.dej2000) WITHIN Circle((189.206250, 62.216111), 0.1)
            AND img.format = 'image/fits' AND img.filter_id = 'W-C-IC'
            AND qso.v_mag < 20
            AND img.region = Box((qso.raj2000, qso.dej2000), 0.02, 0.02)
\end{verbatim}
So it is possible to describe a query to search images for multiple
objects in a single sentence.

\section{DATA SEARCH PROCEDURE} 

We have developed a search engine for federating the VO compliant data services.
The target data services are SIAP/SSAP based image/spectrum data services and
BASIC/FULL skynode catalog services.
The query to those data services is described in the JVOQL-2.
The table join is performed at the JVO portal if the queried services don't 
support the xmatch query.
If there is a service that supports the xmatch query, the xmatch related with the
table belonging to the service is performed at the data server.
SIAP/SSAP services don't accept a SQL query, so the JVOQL-2 is
translated to an HTTP get query.
JVO skynodes are implemented based on the ADQL 1.0 while the other VO skynodes
are still based on the ADQL 0.7.4 as of 2005, so we needed translation
between ADQL 0.7.4.
The namespace of the VOtable is also different for each services, so we
also needed to translate the VOTable to the one with a canonical namespace.

The procedure for a query to multiple databases is as follows:
\begin{enumerate}
   \item Make a list of tables to be queried. For each table, check
	 whether a xmatch query are supported. Check whether a search
	 region is specified. If it is specified, apply it to all the
	 tables which is directory or indirectly cross-matched with the
         table on which the search region is specified.
   \item For each table which does not support a xmatch query and has a 
	 condition associated with the other tables, wait until the
	 query of step 3 and 4 will finish.
   \item For each table which does not support the xmatch query and does
	 not have a condition  associated with the other tables, execute
	 the query for the table.
   \item For each table which supports the xmatch query, estimate a
	 query cost by querying the number of record to be returned.
         Then execute the query to the table which got the lowest cost,
	 and execute xmatch queries to the other tables in accordance
	 with the query cost.
   \item Join the query result of 3 and 4 and execute queries of step 2.
         Finally join all the query results.
\end{enumerate}

At each step of the query, a proper stub is created and used for calling
a service.
Inside the stub, the JVOQL2 is translated to an ADQL for the case of
SkyNode or HTTP URL query string for the case of SIA or SSA service. 

\section{Overall performance test of the JVO}

\begin{figure}
   \begin{center}
   \begin{tabular}{c}
   \includegraphics[width=0.6\textwidth]{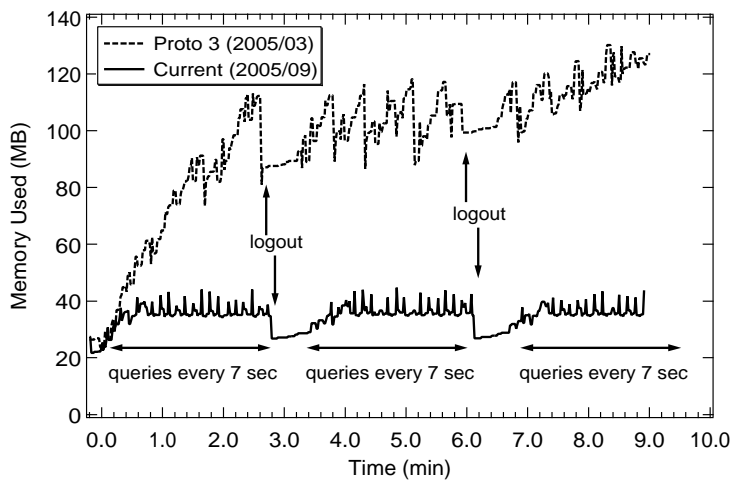}
   \end{tabular}
   \end{center}
   \caption[memory] 
   { \label{fig:memory} 
   Result of the performance test (memory usage history).
   }
\end{figure}

The performance of the JVO portal are measured to make it a robust and
reliable service.
Especially a memory usage is the most critical one to deal with a huge 
number of accesses, and it was expected that, in our AXIS implementation, 
a lot of memory are used for XML to Java object deserialization. 
The size of the XML data is increased by about 50~\% after the
deserialization.
We noticed that proto 3 is consuming a lot of memory and a part of them
are not properly freed.
This is now fixed by making a constraint on the memory usage of each 
user and enforcing an active garbage collection (Fig.~\ref{fig:memory}).
%

We also conducted a simulation of the multiple user accesses using the
Apache JMetter\footnote{http://jakarta.apache.org/jmeter/}.
The simulation is performed under the following condition:
near simultaneous accesses from 100 users;
each user repeats a query 10 times in a hour at a random timing;
each user displays the result for each query.
The major specifications of the portal server are:  
dual AMD Opteron Processor 250 and 16 GB memory.
The memory and cpu usages on the portal server were monitored, and
the response time from the portal was measured on the client.
The error response were also monitored by looking at the HTTP code.
The result of the endurance test was almost satisfactory:
The number of error responses was only 10 for 1000 queries and
the error was caused by the XML DB.
We are requesting the DB developer to fix the problem.
The response time was typically less than 100 ms.
%


\section{WORKFLOW} 

A measurement of the physical properties of an astronomical object is 
usually made by using several independent tools.
In addition, the same measurement will be repeated by changing the 
parameters for the measurement and/or adapting it to the other objects, 
For an example, lets assume the case of measuring the environment of QSO.
In this case, at first, one needs to get a catalog of QSO and also deep 
multi-color images around each QSO.
Next he will do a source detection for each image using a software such
as the SExtractor, and create the source catalog.
To estimate the photometric redshift, he will do a cross-match between
the catalog obtained for different colors and adapt it to the photometric
redshift calculator, such as the HyperZ.
Finally, he will select galaxies which are expected to be at the same
redshift as that of the QSOs, and calculate a cross-correlation between 
the QSO and the galaxies.
Once this flow is described as a workflow, it is easily executed for
different data set, different parameters, and by different analysis tools.
It's ok to write down this kind of flow by a specific language, 
such as shell script, perl, python, and so on.
However, it is better to have an abstract description which does not depends 
on any execution environment, since the script written for a specific environment 
might not work on the other environment.
Workflow is an abstract flow descriptor, and is defined independently from
execution system.
So it is highly re-usable at any environment and suitable for exchange
among people to share the analysis procedure.
The workflow system is recently appearing on the business world for building 
a business logic by re-using the existing web services.
The workflow has been designed for that purpose.
So once the analysis tools are provided through the web service, it can be
easily re-used from the workflow and a user does not need to install the
software on his computing environment.
From such a viewpoint, we developed workflow system to conduct a scientific 
analysis flow~\cite{Tanaka2006}.
The system is divided into two layers.
One is a workflow description language.
Another is a workflow execution system.
They are independent each other so that the workflow can be executed on any 
kinds of execution environments.

\subsection{WORKFLOW DESCRIPTION LANGUAGE}

We defined a workflow description language (WDL) based on the Business Process 
Execution Language for Web Services (BPEL4WS).
BPEL4WS is ``a language for the formal specification of business processes
and business interaction protocols'' designed under the collaboration among
IBM, Microsoft and BEA.
We reduced and extended the BPEL4WS syntax for adapting our purposes.

The WDL is described in XML format, ant the root element is \verb|workflow|.
The root element may have a \verb|variables| element and an activity type element.
The \verb|variables| element defines the name and data type of the variables used
in the document.
The \verb|activity| is an abstract element, it can be one of the sequence activity 
\verb|sequence| and \verb|flow|, basic activity \verb|script|, \verb|command|
and \verb|invoke|, control activity \verb|if|, \verb|switch|, \verb|for|, 
\verb|while|, and \verb|parfor|, or set activity \verb|set|.
\verb|invoke| element is used to describe an execution of a web service.
%
%
For an example, source extractor web service can be invoked with the following \verb|invoke| 
element.
\begin{verbatim}
<workflow>
   <variables>
      <variable name="url" type="String">
         <value>http://jvo.nao.ac.jp/fits/image.fits</value>
      </variable>
      <variable name="votable" type="VOTable"/>
      <variable name="file" type="String">
         <value>result.xml</value>
      </variable>
   </variables>
   <sequence>
      <invoke identifier="ivo://jvo/tools/sextractor" operation="performForURL">
         <input><varRef>url</varRef></input>
         <output><varRef>votable</varRef></output>
      </invoke>
      <command xsi:type="builtin" name="storeVOTable">
         <input>
            <varRef>votable</varRef>
            <varRef>file</varRef>
         </input>
      </command>
   </sequence>
</workflow>
\end{verbatim}
\verb|command| element is used to execute a local command or call a local
Java class method on the workflow execution environment.
Several built-in commands are provided: 
\verb|executeQuery| to execute a data query,
\verb|storeVOTable| to save a VOTable on the disk,
\verb|loadVOTable| to read a VOTable on the disk, and so no.

See Appendix~\ref{sect:wdl_spec} for more details.

\subsection{WORKFLOW EXECUTION SYSTEM}

A workflow is executed by translating the workflow description language to a
Groovy script.
Groovy is a Java based script language and support most of the Java syntax, 
so it can run on the same Java VM as the portal service.
The translation is performed by XSLT, so we just need to write a style sheet.
It is possible to execute the workflow on any other script language, by preparing
the style sheet that translate the XML workflow language to the specific language.
The activity element is replaced by a statement calling an JAVA executer
method, and the element of the loop and the conditional branching are
described by the Groovy script itself.
%

\section{GRID COMPUTING SYSTEM} 

We are going to provide analysis services to the users of JVO portal.
It is especially important to provide server side data reduction system for
data of Subaru SuprimeCam, since the amount of data is very huge and is difficult
to transfer all the data through the internet.
As the SuprimeCam have very distinctive features of wide and deep imaging 
capability, there are a lot of demands to do a survey using the data.
One solution for enabling such a study is to analyze the data at the place 
where it exists and to return only the reduced result.
For an example, if user can execute a source extraction software on the data 
server over the internet, he/she does not need to retrieve all the image to the 
local server.
For providing a computing service to a varying number of users, it is required to 
make a scalable computing system.
The Monitoring and Discovery Service (MDS) is constructed to build a grid computing
system.
Static properties such as a cpu type, number of cpu, memory size, IP address and so on
of each computing server are registered on the MDS at once.
Dynamical properties such as a number of submitted jobs, load average, and so on 
are notified periodically by the computing server to the MDS server.
The interface to the MDS and the analysis service are defined by a Web Service 
Description Language (WSDL), so that a client can invoke the services using a SOAP 
message protocol.
For invoking a asynchronous web service, the MDS is notified a job status by the 
computing server, and a client of the grid system can get the job status by polling 
to the MDS server.
It is also possible to do a polling to the computing server directory, but it is
better to reduce the load on the analysis server.
The asynchronous interface returns a job id immediately after the job submission,
then the job id and the server id are used for querying the job status.
When job status is change from ``running'' to ``finished'', the client request the 
result to the analysis server.
Then the analysis server returns an URL to retrieve the result, and finally the 
client get the result and clear the job on the analysis server. 
%

\section{SCIENCE USE CASE} 

\begin{figure}
   \begin{center}
   \begin{tabular}{c}
   \includegraphics[width=0.8\textwidth]{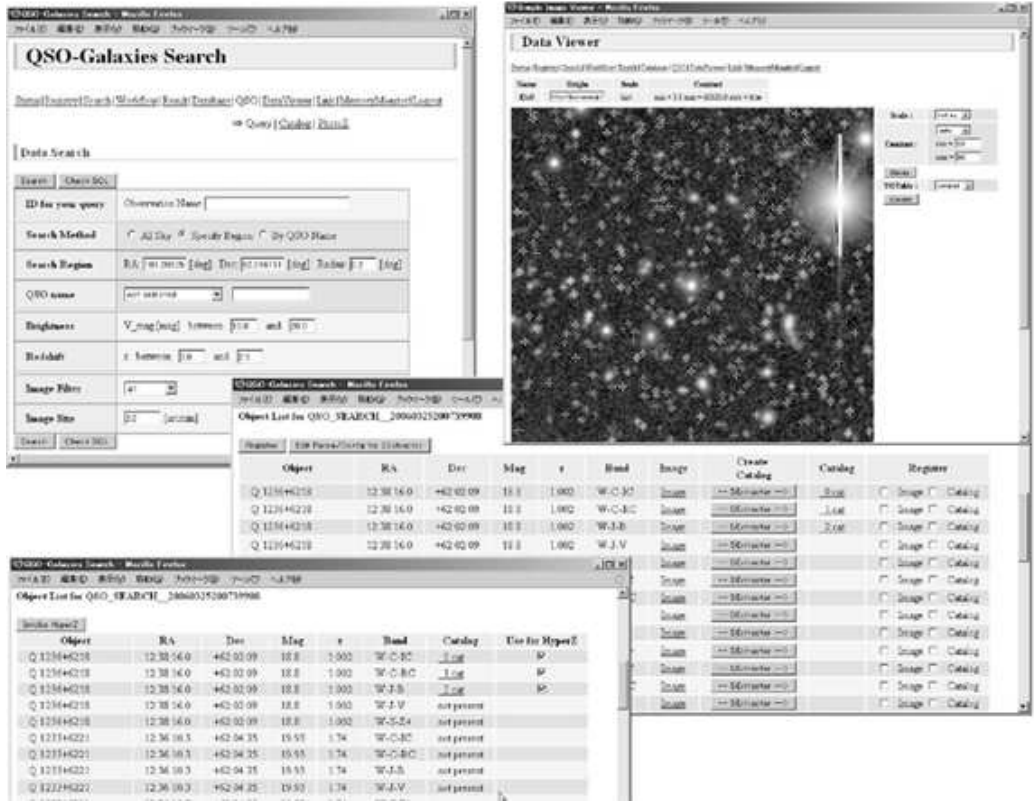}
   \end{tabular}
   \end{center}
   \caption[example] 
   { \label{fig:example} 
      GUI dedicated for the QSO environmental study.
   }
\end{figure} 

We applied the JVO system to a cosmic string search\cite{Shirasaki2004} and
the study of environment of QSOs.
Here we will discuss the later use case.
QSO environment study was made by combing the QSO catalog service and
the Subaru image data service. 
Since we don't have a reduced data archive yet, raw data of pre-selected 
five fields are retrieved from the SMOKA and MASTARS service operated by
NAOJ, and they are reduced with a standard analysis tool and registered 
as a skynode database. 
The fields are selected through cross-matching between QSO database and 
Suprime-Cam frame database. 
A workflow for this study is as follows:
1. Select QSO coordinates from the QSO database.
2. Search multi-bands imaging data which covers the QSO regions.
3. Create a catalog from the imaging data by invoking a SExtractor Web service.
4. Estimate the objects' photoZ around the QSO.
5. Clustering Analysis.
The interactive GUI was prepared for this specific study as shown in
Fig.~ref{fig:qso}.
On this GUI one can specify the search criteria for the properties of
the QSO and retrieved image and execute the query.
At the page of the query result, there are buttons to invoke the
sextractor web service for each retrieved image.
The result of the source extractor can be overlaid on the image.
It is also possible to invoke the hyperz web service which calculate the
photometric redshift from the multi-color catalog.
The above workflow were also executed on the JVO workflow system, and we
have successfully completed the workflow.
%
%
%

\section{FUTURE PLAN}

The JVO portal service will partially become public soon in this year
2006 or early of 2007.
More user friendly interface for query, data visualization, workflow
editor will be provided in near future.
We will collaborate with the the Strasbourg astronomical Data Center
(CDS) and Astro Grid project for development of the data visualization
and workflow system.
We will also start to provide data of reduced data of Subaru Suprime-Cam 
on the VO standard service in 2006.
%

\appendix    

\section{WORKFLOW DESCRIPTION LANGUAGE SPECIFICATIONS} 
\label{sect:wdl_spec}

\subsubsection*{\mbox{Workflow}}

A root node of the XML representation of a workflow is a \verb|Workflow|
element.
The \verb|Workflow| element has exactly one \verb|name| element and \verb|author|
element.
The \verb|Workflow| element may have optional elements \verb|identifier|,
\verb|create_date|, \verb|update_date|, \verb|description|, \verb|status|, 
\verb|variables| and an activity type element.
What user or GUI should specify are the \verb|variables| and activity type element.
The \verb|name|, \verb|author| and \verb|description| may be specified by the user,
and if not specified workflow engine fill the \verb|name| and \verb|author| elements.
Currently 11 activity types are defined: 
\verb|sequence| and \verb|flow| elements for describing multiple activities,
\verb|invoke|, \verb|command| and \verb|script| elements for executing a job,
\verb|set| element for assigning data to a variable,
\verb|for|, \verb|parfor| and \verb|while| elements for repeating an activity,
\verb|if| and \verb|switch| elements for selecting one branch of activity according
to the specified criteria.

\subsubsection*{variables and variable}

The \verb|variables| element may have an unlimited number of a
\verb|variable| element.
The \verb|variable| element defines the name and data type of a variable used in the
workflow.
The name of the variable is defined at the \verb|name| attribute and the data type is
defined at the \verb|type| attribute.
An initial value may be specified by \verb|value| element under the \verb|variable| element.
The variable name must be unique in a workflow document.
The data types that can be specified are \verb|int|, \verb|double|, \verb|String|,
\verb|VOTABLE|, \verb|DataHandler|, \verb|List| and \verb|Map|.
Array types of these data type (\verb|int[]|, \verb|double[]|, ...) are
also available.
\verb|DataHandler| is used to transfer a file as an attachment of a SOAP message.
\verb|List| and \verb|Map| are the same as the Java objects of \verb|java.uti.List| and 
\verb|java.util.Map|.

\subsubsection*{sequence, flow}

The \verb|sequence| and \verb|flow| elements may have unlimited number
of multiple  activity type elements.
The activities represented by child elements of the \verb|sequence| element are 
executed in  sequence, while those of the \verb|flow| element are executed in 
parallel.

\subsubsection*{invoke, command, script}

The \verb|invoke|, \verb|command| and \verb|script| elements are used to
execute an unit of job.
All of these element have a common attribute \verb|activity_name|, and may have
a child element \verb|activity_status|, \verb|input| and \verb|output|.

\verb|invoke| element represents an activity which invokes a web service.
The \verb|invoke| element has attributes \verb|identifier|, \verb|porttype|, 
\verb|namespace|, \verb|url| and \verb|operation|, and may have an unlimited number 
of \verb|input| child elements and one child element \verb|output|.
To uniquely specify an web service operation, a service identifier, a port type name, 
a namespace of the port type and an operation name are required to be specified at the
attribute of the \verb|invoke| element.
In most of the cases, only the \verb|identifier| and \verb|operation| name are
enough to pinpoint the operation, so the attributes that must be supplied are
\verb|identifier| and \verb|operation|.
The other attributes may be omitted if there is no multiple operation for the
set of \verb|identifier| and \verb|operation|.
The \verb|url| is resolved by the workflow engine, so it does not need to be specified
by the user.
The child element \verb|input| specify the input parameter that is passed to the operation, 
and the \verb|output| child element specify a variable that receive the returned value of 
the operation.

The \verb|command| represents an activity which executes a local command on the workflow
execution environment.
The \verb|command| element has an attribute \verb|xsi:type|, which must have a value
either of \verb|builtin| or \verb|classMethod|.
When \verb|builtin| is specified, \verb|name| attribute must be suppled to the \verb|command|
element and it specify the what built-in command is executed.
Currently defined built-in commands are: \verb|executeQuery|, \verb|invoke|, \verb|storeVOtable|,
\verb|loadVOTable| and \verb|laodAsDataHandler|.

The \verb|script| element represents an activity which execute an groovy script specified 
at the \verb|input| element.

\subsubsection*{set}

The \verb|set| element represents an activity which sets a value to the specified variable.
An attribute \verb|name| must be given, which specifies the variable that is set a value 
by this activity.
There are three way to set a value.
One is to set the value by specifying a literal expression at a \verb|literal| attribute.
The variable value can be set by specifying the variable name at a \verb|ref| attribute.
The value of a property of an object type variable can be set by
specifying the property name and the variable name at the \verb|property|
and \verb|ref| attributes, respectively.
To set a value contained in a \verb|List| type variable, an expression like \verb|param[n]| 
can be used.
To set a value contained in a \verb|Map| type variable, the name of a string parameter 
corresponding to key name and of a \verb|Map| type  variable should be specified at 
\verb|property| and \verb|ref| attribute, respectively.

\subsubsection*{for, parfor, while}

The \verb|for| element represents an activity which repeats an activity specified in a child
element to the extent specified at the attributes.
There are two methods to specify the number of repetitions.
One is to specify the collection type variable (List, Map and Array) at the \verb|items| 
attribute.
The child activity is repeated by a number of elements of the collection type variable.
The number of execution is refereed by a \verb|_count| variable, and the element of the
collection type variable \verb|param| is refereed by \verb|param[_count]|.
Another is to specify the \verb|init|, \verb|max| and \verb|step|.
This is the same as the for loop of C-language and Java.

The \verb|parfor| element represents an activity which does the almost same thing as the 
\verb|for|, but it execute an child activity in parallel.

The \verb|while| element represents an activity which repeat an child activity while
the condition specified at the \verb|condition| attribute is true.

\subsubsection*{if, switch}

The \verb|if| element represents an activity which execute one branch of the 
child activity according to the condition specified at the \verb|condition| 
attribute.
If the condition is evaluated as true, child activity specified by the child element 
\verb|then| is executed. 
Otherwise, child activity specified by the child element \verb|else| is executed if
the \verb|else| element exists.

The \verb|switch| element represents an activity which execute one branch of child 
activity according to the value specified at the \verb|test| attribute.
The \verb|switch| element have an unlimited number of child elements of \verb|case|,
and may have a child element of \verb|otherwise|.
The \verb|case| element has an attribute \verb|condition|, where the condition that
the child activity is executed is specified.
If any child activity of \verb|case| element is not executed and there is the
\verb|otherwise| element, the child activity of the \verb|otherwise| element is
executed.

\acknowledgments     

This work was supported by the JSPS Core-to-Core Program,
Grant-in-aid for Information Science (15017289 and 16016292)
and Young Scientists (B) (17700085) carried out by the Ministry of
Education, Culture, Sports, Science and Technology of Japan.


\bibliography{yshirasa}   
\bibliographystyle{spiebib}   

\end{document}